\documentclass[%
 reprint,
 amsmath,amssymb,
 aps,
]{revtex4-2}

\usepackage{graphicx}%
\usepackage{dcolumn}%
\usepackage{bm}%
\usepackage{lineno}

\usepackage{lineno}
\usepackage{makecell}
\usepackage{textcomp}
\usepackage{subfigure}
\usepackage{threeparttable} 
\graphicspath{{figures/}}
\usepackage{xcolor}
\usepackage{booktabs}

\begin{document}

\preprint{APS/123-QED}

\title{Room-temperature waveguide-coupled silicon single-photon avalanche diodes}

\author{Alperen Govdeli$^{1,2}$}
\email{alperen.govdeli@mail.utoronto.ca}
\author{John N. Straguzzi$^{1}$}%
\author{Zheng Yong$^{2}$}%
\author{Yiding Lin$^{1}$}%
\author{Xianshu Luo$^{3}$}
\author{Hongyao Chua$^{3}$}
\author{Guo-Qiang Lo$^{3}$}
\author{Wesley D. Sacher$^{1}$}
\author{Joyce K. S. Poon$^{1,2}$}
\email{joyce.poon@mpi-halle.mpg.de}
\affiliation{%
 $^1$Max Planck Institute of Microstructure Physics, Weinberg 2, 06120 Halle, Germany \\
 $^2$Department of Electrical and Computer Engineering, University of Toronto, 10 King's College Road, Toronto, Ontario, M5S 3G4, Canada\\
 $^3$Advanced Micro Foundry Pte Ltd, 11 Science Park Road, Singapore Science Park II, 117685, Singapore, Singapore\\
}%

\date{\today}

\begin{abstract}
Single photon detection is important for a wide range of low-light applications, including quantum information processing, spectroscopy, and light detection and ranging (LiDAR). A key challenge in these applications has been to integrate single-photon detection capability into photonic circuits for the realization of complex photonic microsystems.  Short-wavelength ($\lambda < 1.1$ {\textmu}m) integrated photonics platforms that use silicon (Si) as photodetectors offer the opportunity to achieve single-photon avalanche diodes (SPADs) that operate at or near room temperature.
Here, we report the first waveguide-coupled Si SPAD. The device is monolithically integrated in a Si photonic platform and operates in the visible spectrum. The device exhibited a single photon detection efficiency of $>$ 6$\%$ for wavelengths of 488 nm and 532 nm with an excess voltage less than 20$\%$ of the breakdown voltage. The dark count rate was below 100 kHz at room temperature, with the possibility of improving by approximately 35$\%$ by reducing the temperature to -5$^{\circ}$C.
\end{abstract}

\maketitle

\section{INTRODUCTION}

Low-level light or single-photon detection is important in applications such as quantum information processing \cite{niffenegger2020integrated, mehta2020integrated, hadfield2009single, ivory2021integrated, setzer2021fluorescence, kwon2023multi}, fluorescence microscopy and spectroscopy \cite{bruschini2019single, michalet2014silicon}, and light detection and ranging (LiDAR) \cite{yu2017fully, buller2007ranging}. Fundamental scientific research in these domains is progressively transitioning toward the development of integrated microsystems, to facilitate more practical applications in the field and support increasingly complex experiments that require a large number of channels \cite{dong2022high, palm2023modular, puckett2021422, edinger2021silicon, bogaerts2020programmable, wan2020large, arrazola2021quantum, madsen2022quantum}. This transition motivates the integration of single-photon detectors onto photonic integrated circuits (PICs), especially those based on silicon (Si) photonics, for which the microelectronics fabrication processes and foundry manufacturing ecosystem enable very large-scale integration on 200mm or 300mm diameter wafers. Technologies for single-photon detection include superconducting nanowire single-photon detectors (SNSPDs), transition edge sensors (TESs), and single-photon avalanche diodes (SPADs).  
SNSPDs and TES use superconductors, boast low dark counts, and work broadly for light with wavelengths spanning from the visible (VIS) to near-infrared (NIR) spectrum.  
SNSPDs have been integrated on Si \cite{pernice2012high} and silicon nitride (SiN) \cite{schuck2016quantum} waveguides, where they achieve high photon detection efficiency (PDE) ($>$ 90$\%$) with low dark count rates (DCRs) of $<$ 100 Hz and minimal jitter $<$ 20 ps \cite{pernice2012high}.  
TES that are not integrated with waveguides have high PDEs near 81$\%$ \cite{fukuda2009photon} in the near-infrared (NIR) range ($\lambda \sim 850$ nm), but have relatively poor timing jitter ($\sim$100 ns \cite{fukuda2009photon}) when compared to SNSPDs. When integrated with waveguides, the TES have exhibited a lower PDE ($< 10\%$) \cite{gerrits2011chip}. A major drawback of superconductor-based devices is that they operate at cryogenic temperatures ($<$ 5 K). 
On the other hand, SPADs, which are PN or PIN junction diodes, usually operate at or near room temperature. In Geiger mode, where the photodetector is biased well above the breakdown voltage ($|V_{bias}| > |V_{br}|$), incoming photons generate free charge carriers to cause high avalanche currents under a high electric field in the junction.

For wavelengths shorter than 1100 nm, in the VIS and NIR parts of the spectrum, Si SPADs are common and are often monolithically integrated with complementary metal-oxide-semiconductor (CMOS) electronics. There are various VIS spectrum SPADs realized in standard CMOS \cite{lee2015first,niclass2007single,panglosse2021modeling,charbon2013geiger,veerappan2014substrate}. For example, CMOS SPADs with a P+/N-well junction operating in VIS with moderate PDEs (41$\%$ at $\lambda=$ 450 nm \cite{niclass2007single}, $\sim$48\% at $\lambda=$ 480 nm \cite{veerappan2014substrate}) and DCRs of the order of kHz (800 kHz in \cite{niclass2007single}, 12.84 Hz/$\mu$m$^2$ in \cite{veerappan2014substrate}) at room temperature have been shown.      
In these structures, the PN junction is vertical. It is isolated from the Si substrate and is surrounded by a doped guard-ring. These features improve light absorption, protect the junction from high electric fields, and isolate charge carrier drift in the substrate \cite{lee2015first,niclass2007single,veerappan2014substrate}. 
However, monolithic integration of conventional CMOS SPADs, primarily for free-space visible-light detection, with planar nanophotonic waveguides, is challenging, requiring efficient optical coupling and compatible geometries. In the telecom regime, waveguide-coupled SPADs using Ge-on-Si, where Ge is the light-absorbing layer, have been demonstrated \cite{martinez2017single}.  The reported PDE was $\sim$5$\%$ at $\lambda=$ 1310 nm, and the DCR was 534 kHz, but the operating temperature was 80 K \cite{martinez2017single}.

In this work, we report, to the best of our knowledge, the first waveguide-coupled Si SPAD. The device is suitable for detection of VIS light. The preliminary results of the device were reported in \cite{govdeli2023monolithically}. The device was monolithically integrated within a Si photonics platform, featuring various photonic components such as low-loss VIS spectrum waveguides \cite{sacher2019visible}, edge couplers \cite{lin2021low}, thermo-optic phase shifters \cite{yong2022power}, and electrothermally actuated micro-electromechanical systems (MEMS) cantilever beam scanners \cite{sharif2023microcantilever}. We have previously reported its linear mode operation in \cite{lin2022monolithically}. Here, we extend our work to demonstrate single-photon (Geiger-mode) operation. We characterize the photon detection efficiency and dark count rate at room temperature. This work opens the path toward the integration of single-photon detectors in very large-scale integrated photonic circuits operating near room temperature \cite{arrazola2021quantum, madsen2022quantum, bartolucci2023fusion, dong2022piezo, wan2020large, kim2020hybrid}.

\section{RESULTS}
\subsection*{Device design}

Figure \ref{fig:LinearMode_PD_CrossSection}(a) shows a schematic of the cross-section of the waveguide-coupled Si SPAD. Details of the device geometry are reported in \cite{lin2022monolithically}. The structure is composed of a Si mesa with a PN junction under a SiN waveguide. Input light in the SiN waveguide evanescently coupled into the Si mesa. The routing SiN waveguides were designed to have a thickness of 150 nm and a width of 500 nm. Adiabatically tapered edge couplers were used to facilitate fiber-to-chip coupling. The SiN width reduced from 5.2 {\textmu}m at the facet to 500 nm, the nominal width of the waveguide ($W_{gw}$) in the chip.  
Figure \ref{fig:LinearMode_PD_CrossSection}(b) and \ref{fig:LinearMode_PD_CrossSection}(d) show the top-view optical micrograph and schematic of the SPAD. The waveguide width was narrowed to $250\pm 25$ nm ($W_{gn}$) along the length of the junction at the Si mesa facet to enhance the coupling efficiency between the SiN waveguide and the PN junction. The interlayer separation was about 150 nm. 

\subsection*{Linear mode operation}

\begin{figure*}
    \centering
     \includegraphics[width= \textwidth]{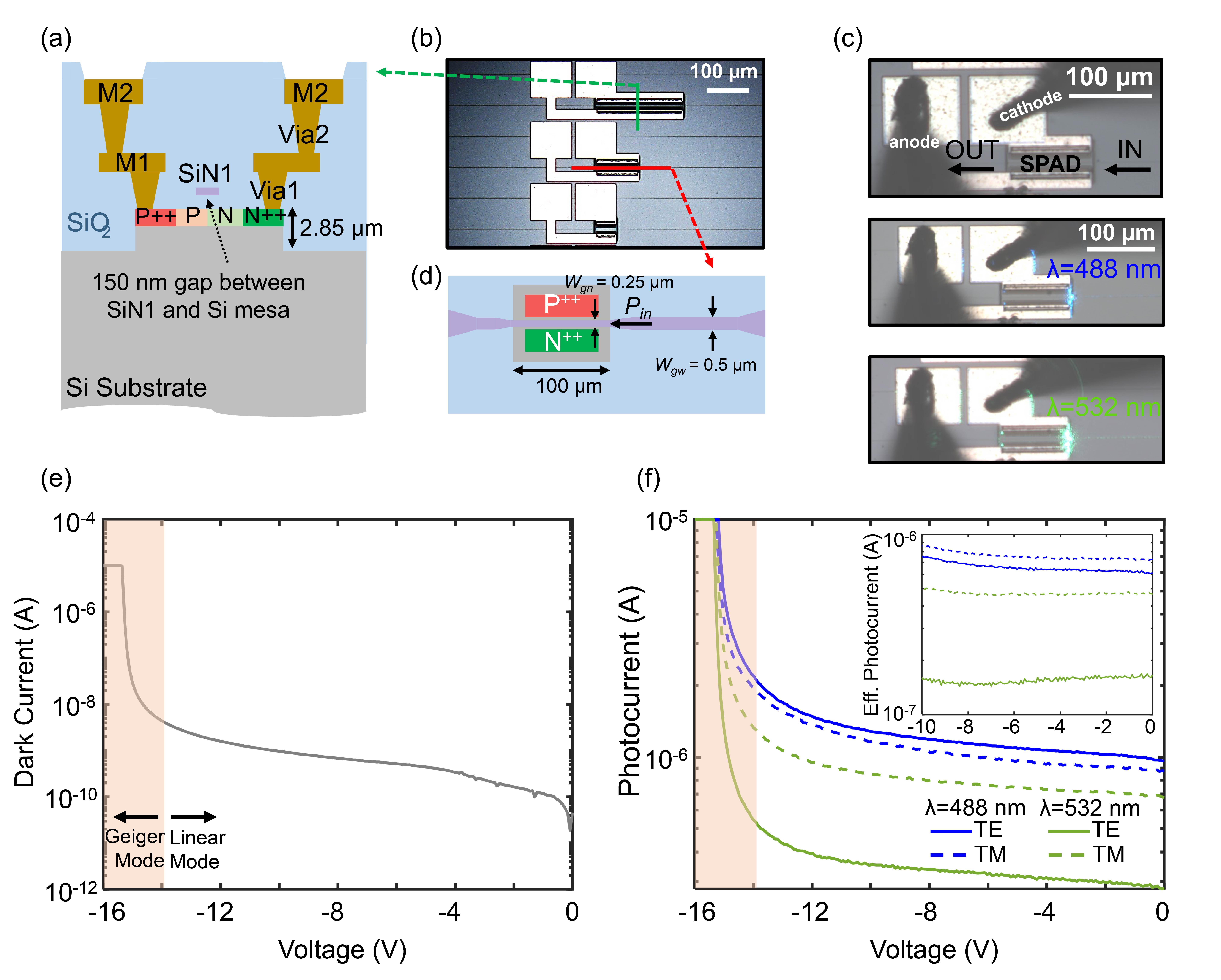}
    \caption{Waveguide-coupled visible-light SPAD: (a) Schematic of the cross-section. (b) An optical micrograph of detectors with different junction lengths, and (c) the SPAD characterization performed at $\lambda=$ 488 nm and 532 nm. The device was contacted by DC electrical probes connected to the anode and cathode metal pads of the SPAD. (d) Top-view schematic of the SPAD structure with a junction length of 100 {\textmu}m. (e) Current-voltage (I-V) characteristics of the SPAD in linear mode (avalanche mode) in the absence of incoming light (i.e., dark current) and reverse bias at room temperature. (f) I-V characteristics of the SPAD in linear-mode operation under light illumination at 488 nm (blue, $P_{in, TE}=2.13$ $\mu$W, $P_{in, TM}=2.33$ {\textmu}W) and at 532 nm (green, $P_{in, TE}=0.85$ $\mu$W, $P_{in, TM}=1.85$ {\textmu}W), where solid and dashed lines are for TE- and TM-polarized light, respectively. Inset: effective photocurrent due to the incoming light without the contribution of the stray light. The shaded area in (e) and (f) represents the Geiger mode operation region for the SPAD.}
    \label{fig:LinearMode_PD_CrossSection}
\end{figure*}

In the linear mode operation of the device, the number of generated charge carriers, $n_e$, is linearly proportional to the number of absorbed photons, $n_p$, as given by 
\begin{equation}\label{eq:ChargedCarriers}
    n_{e} \propto \eta \cdot  M \cdot n_p,
\end{equation}
where $\eta$ and $M$ are internal quantum efficiency and avalanche multiplication factor, respectively. The measured current-voltage (I-V) characteristics in Fig. \ref{fig:LinearMode_PD_CrossSection} (f) show the linear-mode operation of the device at $\lambda=$488 nm and $\lambda=$532 nm for transverse electric (TE) and transverse magnetic (TM) polarized input light. The results are consistent with the typical avalanche photodetector (APD) reported in \cite{lin2022monolithically}. A photocurrent, $I_{eph}$, of 0.68 $\mu$A (0.76 $\mu$A) was observed for a TE (TM)-polarized light with an optical power, $P_{in}$, of 2.13 $\mu$W (2.33 $\mu$W) at $\lambda=$488 nm with a reverse bias of 8 V. Measurements were also repeated for $\lambda=$532 nm. The corresponding responsivity and external quantum efficiency (EQE) were 0.32 A/W (0.17 A/W) and 81\% (40\%) for TE-polarized light at $\lambda=$488 nm ($\lambda=$532 nm), respectively. The linear mode performance of the device is summarized in Table \ref{table:LinearMode}. The measured photocurrent includes the absorption of stray light in the cladding, which was not coupled to the waveguide. Thus, the effective photocurrent ($I_{eph}$, Fig. \ref{fig:LinearMode_PD_CrossSection}(f)-inset) due to the light coupled to the waveguide was measured by removing the contribution of the stray light (see the Methods section). The dark current, as depicted in Fig. \ref{fig:LinearMode_PD_CrossSection} (e), was 0.16 nA at a reverse bias of 2 V and room temperature.

\begin{table}[]
\caption{Summary of the APD performance in linear mode at room temperature}
\label{table:LinearMode}
\resizebox{\columnwidth}{!}{%
\begin{tabular}{cccc}
\hline
\textbf{$\lambda$ (nm)} & \textbf{Polarization} &  \textbf{Responsivity (A/W)} & \textbf{EQE (\%)} \\ \hline
488         & TE           & 0.32    & 81       \\
488         & TM           & 0.33    & 83       \\
532         & TE           & 0.17    & 40       \\
532         & TM           & 0.25    & 59       \\ \hline
\end{tabular}%
}
\end{table}

\subsection*{Single photon detection}\label{experimental_procedure}

\begin{figure*}
    \centering
       \includegraphics[width= \textwidth]{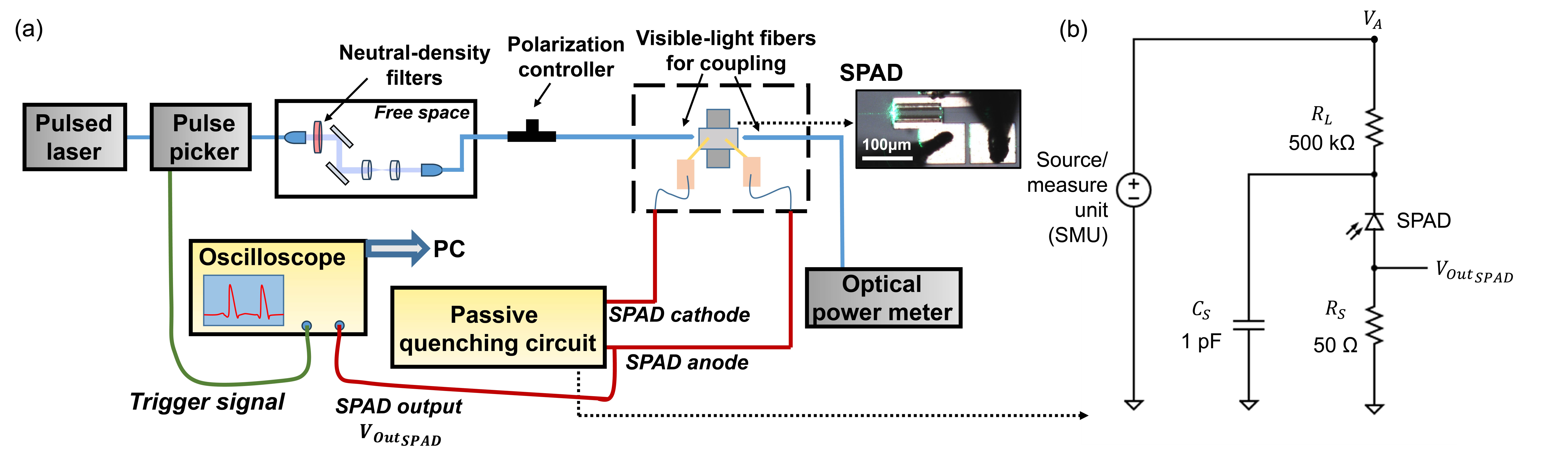}
    \caption{Characterization of the SPAD in Geiger mode: (a) Schematic of the SPAD characterization setup. Blue lines represent optical paths. Red and green lines illustrate the electrical connections. (b) Circuit diagram of the passive quenching circuit used in the measurements.}
    \label{fig:ExperimentSetupSchematic}
\end{figure*}

To characterize the single-photon detection properties of the waveguide SPADs, we used the setup as shown in Fig. \ref{fig:ExperimentSetupSchematic}. The $\lambda=488$ nm or $532$ nm output of a supercontinuum laser source (NKT Photonics SuperK Fianium) was filtered with a spectral bandwidth less than 2.5 nm. The laser repetition rate was 78 MHz, and the full-width at half-maximum (FWHM) pulse width was 10 ps. A pulse picker with an extinction ratio $> 50$ dB reduced the repetition rate to 513 kHz. The light was then attenuated by neutral-density filters to meet the single-photon condition (see Eq.\ref{eq:averagenumber} in the Methods section) and then coupled to the chip via a single-mode visible-light fiber. The chip was enclosed by a black box during the measurements to eliminate the impact of ambient light on the device characterization. We used a simple passive quenching circuit as illustrated in Fig. \ref{fig:ExperimentSetupSchematic}(b) for the SPAD device (details in the Methods section). The output voltage of the device, triggered by the pulse picker, was captured on a 500 MHz digital storage oscilloscope (DSO, Keysight InfiniiVision DSOX4054A) with a data sampling rate of 5 GSample/s. To accumulate sufficient data for statistical analysis, 0.1 ms-long time traces were repeatedly acquired until a total of 25000 pulses, each containing $<1$ photon, had been applied. 

\begin{figure*}
    \centering
         \includegraphics[width= \textwidth]{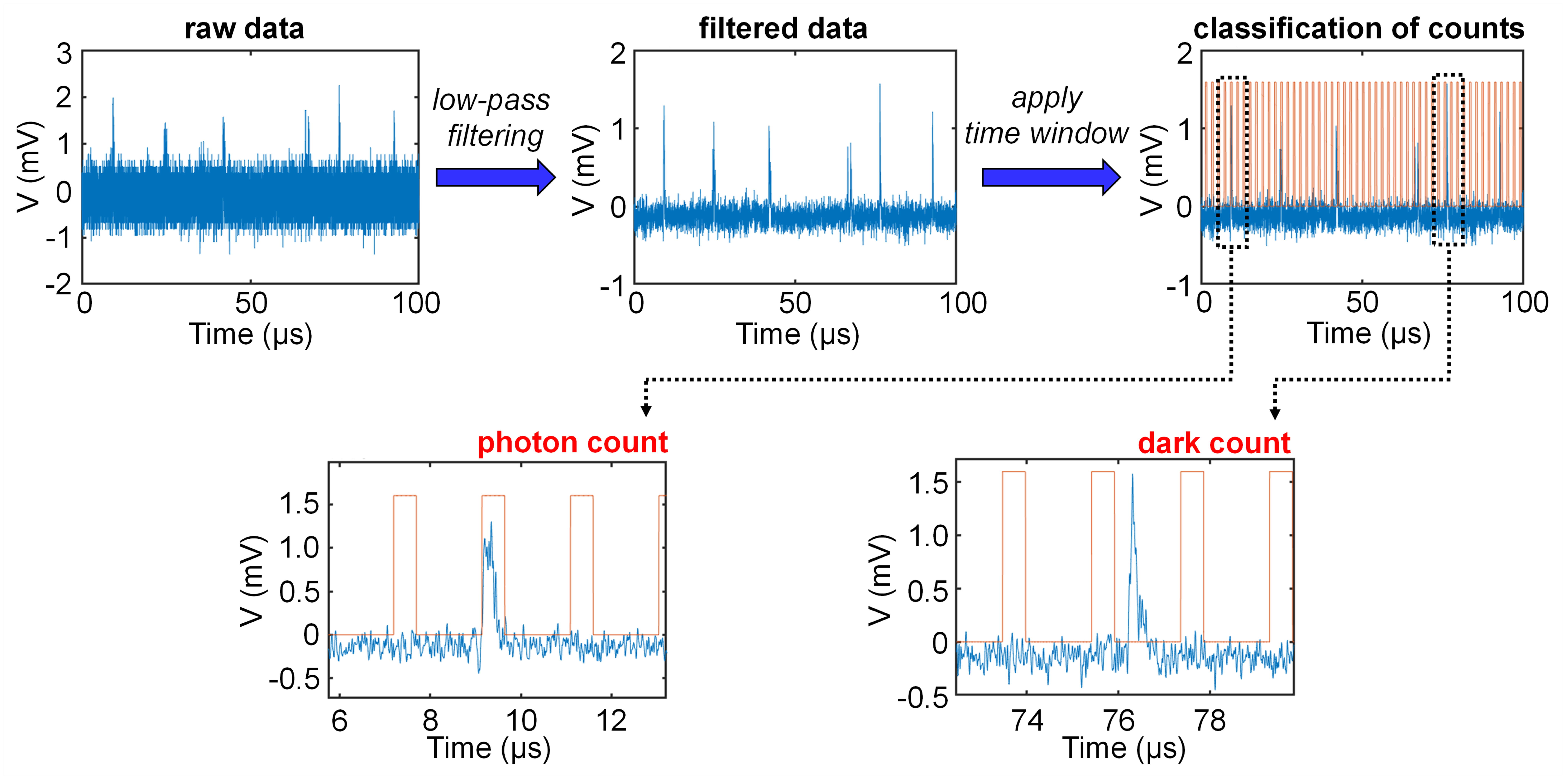}
    \caption{Flow chart of the procedure for the photon and dark count calculations. Time traces of the SPAD output signal were acquired over a duration of 0.1 ms for 500 times for each set of measurements. A total of 25000 light pulses, each with $<$ 1 photon per pulse, were applied. The output signals were low-pass filtered ($<$ 3 MHz) and inserted into time windows, each spanning 500 ns. The classification of individual avalanche pulses as photon and dark counts was determined based on their temporal position within these windows.} 
    \label{fig:calculationflowchart}
\end{figure*}

We analyzed the measurements using the workflow in Fig. \ref{fig:calculationflowchart}. The time traces were low-pass filtered with a cut-off frequency of 3 MHz to suppress high frequency noise from the DSO and measurement setup. Subsequently, time windows were applied to the data for photon counting. Each window, with a width of 500 ns, was centered on the arrival time of the peak of an input pulse. This windowing procedure enabled us to sort the measured counts into photon ($n_{photo}$) and dark ($n_{dark}$) counts, which were then used to calculate the photon detection efficiency (PDE) and dark count rate (DCR), as explained in the Methods section.

\begin{figure*}
    \centering
   \includegraphics[width= \textwidth]{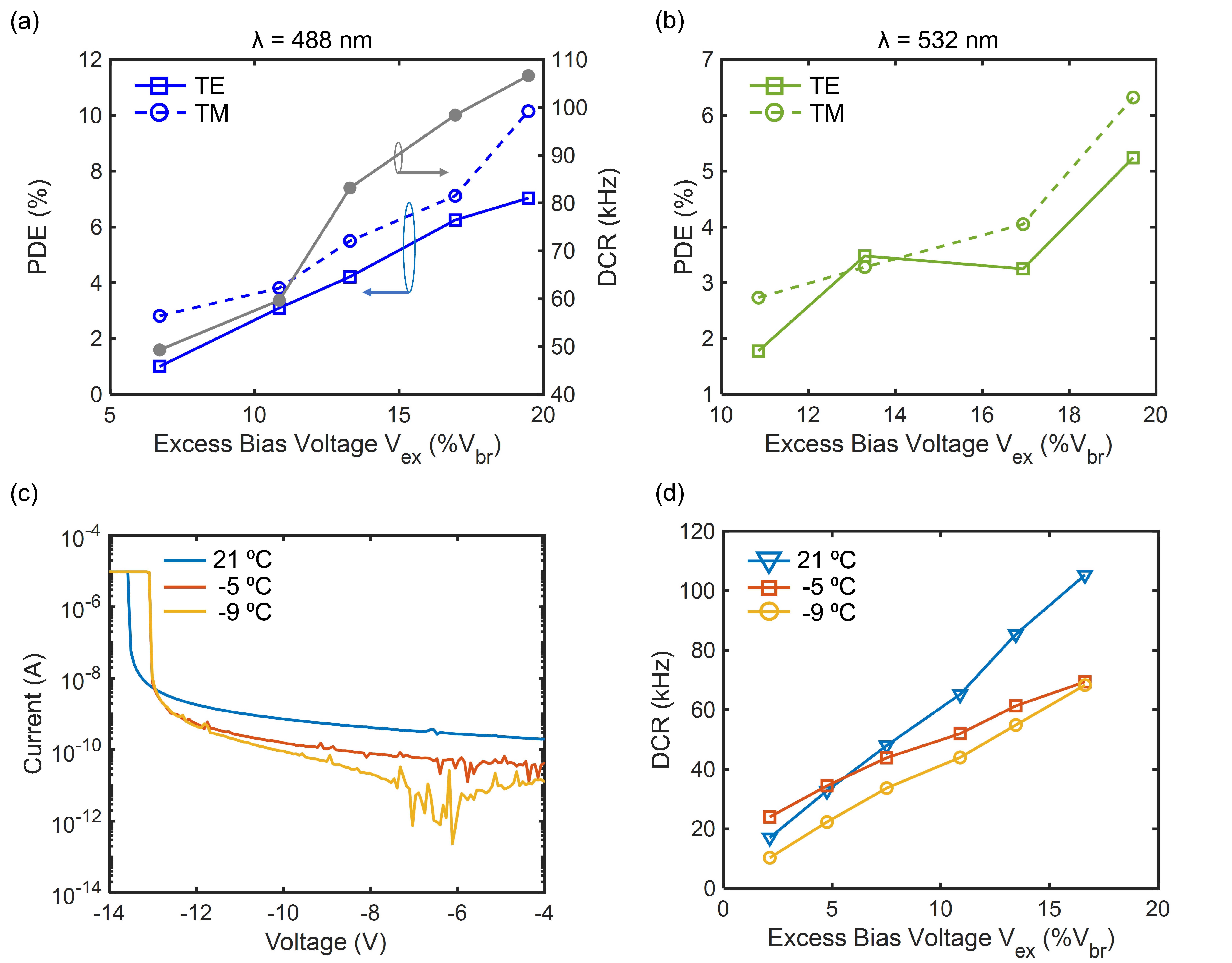}
    \caption{Geiger mode operation of the SPAD. (a) Single photon detection efficiency (PDE) of the SPAD for TE (solid line, square) and TM (dashed line, circle) polarized light at $\lambda=$ 488 nm, and dark count rate (DCR) for an excess bias voltage ($V_{ex}$) range spanning 5$\%$-20$\%$ of the breakdown voltage ($V_{br}$) at room temperature (21$^{\circ}$C). (b) The PDE of the SPAD at $\lambda=$ 532 nm at 21$^{\circ}$C for different excess voltages. (c) Current-voltage characteristics in the absence of incoming light, and (d) dark count rate (DCR) of the SPAD at 21$^{\circ}$C, -5$^{\circ}$C, and -9$^{\circ}$C.}
    \label{fig:PDE_DCR_RoomTemp}
\end{figure*}

Figures \ref{fig:PDE_DCR_RoomTemp}(a) and \ref{fig:PDE_DCR_RoomTemp}(b) show the PDE of the SPAD with a lateral junction length of 100 {\textmu}m measured at various excess bias voltages. The PDE was inferred from the measured avalanche pulses and the number of photons in the waveguide (see the Methods section). These measurements were carried out at room temperature for TE- and TM-polarized light at $\lambda=$488 nm [Fig. \ref{fig:PDE_DCR_RoomTemp}(a)] and 532 nm [Fig. \ref{fig:PDE_DCR_RoomTemp}(b)]. To ensure stable biasing and eliminate the impact of minor voltage drifts, a constant current source ($I_{bias}$) was used for bias.  
The excess bias voltages, $V_{ex}$, were calculated as a percentage of the breakdown voltage, $V_{br}$, using $100\times(V_A-|V_{br}|)/|V_{br}|$, where $V_A$ is the bias voltage.

At $\lambda=488$ nm, PDEs of 4.2$\%$ and 5.5$\%$ were observed with $V_{ex}$ of 13.3$\%$ ($I_{bias}=2$ {\textmu}A) for TE and TM polarization, respectively. The SPAD under the same bias provided a PDE of approximately 3.5$\%$ at 532 nm. As shown in Fig. \ref{fig:PDE_DCR_RoomTemp}, the PDE reached 10$\%$ (6.3 $\%$) at 488 nm (532 nm) with $V_{ex}$ of 19.5$\%$ and $I_{bias}=3\mu$A. 
The further increase in $V_{ex}$ led to a higher PDE since the probability of avalanche, $P_a$, was higher \cite{oldham1972triggering}. However, this and the electric field dependent carrier generation also increased the DCR  \cite{martin1981electric}. As $P_a$ approaches unity with an increase in $V_{ex}$, the PDE reaches its maximum; however, the high electric field still contributes to the growth of DCR. Consequently, $V_{ex}$ is restricted by the signal-to-noise (SNR) ratio.

Figure \ref{fig:PDE_DCR_RoomTemp}(a) shows the DCR as a function of the excess bias voltage of the SPAD at room temperature (21$^{\circ}$C). The DCR was between 49 kHz and 107 kHz with $V_{ex}$ in the range of 6.7$\%$ to 19.5$\%$, and it increased with bias voltage, consistent with the trap-assisted-tunnelling component of Shockley-Read-Hall model \cite{kindt1998modelling}. The impact of thermal generation on DCR was investigated by conducting measurements on a nominally identical device from another chip at lower temperatures. 
The I-V characteristics of the SPAD in linear-mode at 21$^{\circ}$C, -5$^{\circ}$C, and -9$^{\circ}$C is shown in Fig. \ref{fig:PDE_DCR_RoomTemp}(c).
As the temperature decreased from 21$^{\circ}$C to -9$^{\circ}$C, the dark current decreased from 0.4 nA to 20 pA at 8 V reverse bias, and $V_{br}$ shifted from 12.94 V to 12.6 V. 
Similar to the linear-mode dark current, the DCR of the SPAD in Geiger-mode decreased with temperature due to the reduced thermal generation rate in the junction, as shown in Fig. \ref{fig:PDE_DCR_RoomTemp}(d). 
For example, under high electric fields, i.e., $V_{ex}=17\%$, the DCR at room temperature was about 105 kHz, while it reduced by about $35\%$ to about 69 kHz and 68 kHz at -5$^{\circ}$C and -9$^{\circ}$C, respectively.  
Due to its weak dependence on temperature, DCR values at -5$^{\circ}$C and -9$^{\circ}$C were similar under a high electric field, and the effectiveness of further cooling on the DCR was limited. Table \ref{table:GeigerMode} summarizes the performance of waveguide-coupled SPAD in Geiger mode under $I_{bias}$ of 1.5, 2, and 3 $\mu$A at room temperature.

\begin{table}[]
\caption{Summary of the SPAD performance in Geiger mode at room temperature}
\label{table:GeigerMode}
\resizebox{\columnwidth}{!}{%
\begin{tabular}{cc|cccc|c}
\hline
          &         & \multicolumn{4}{c|}{\textbf{PDE (\%)}}                     & \textbf{DCR}   \\
\textbf{$I_{bias}$ ($\mu$A)} & \textbf{$V_{ex}$(\%)} & \makecell{488 nm\\ TE} & \makecell{488 nm\\ TM} & \makecell{532 nm\\ TE} & \makecell{532 nm\\ TM} & (kHz) \\ \hline
1.5       & 10.8    & 3.1        & 3.8        & 1.8        & 2.7        & 60    \\
2         & 13.3    & 4.2        & 5.5        & 3.5        & 3.3        & 83    \\
3         & 19.5    & 7          & 10.2       & 5.2        & 6.3        & 107   \\ \hline
\end{tabular}%
}
\end{table}  %
\section{Discussion}

In this study, we have demonstrated waveguide-coupled visible-light SPADs with PDEs ranging from 5$\%$ to 10$\%$ and DCRs below 100 kHz at room temperature. Notably, higher efficiencies were observed for TM-polarized light due to the increased evanescent coupling efficiency between SiN and Si mesa. As the wavelength increased to 532 nm, the PDE values decreased, which is attributed to the external quantum efficiency of the structure. Although the longer wavelength single-photon detection is beyond the scope of the present work, it is reasonable to assume, based on its linear mode performance (Fig. \ref{fig:LinearMode_PD_CrossSection}(f) and \cite{lin2022monolithically}), that the waveguide SPAD is expected to continue to be effective in Geiger mode. The PDE was highest at around 488 nm, consistent with the maximum external quantum efficiency of the APD \cite{lin2022monolithically}. The PDE can be improved by increasing the external quantum efficiency through narrowing the SiN waveguide for wavelengths closer to 400 nm and choosing an optimal separation between the SiN and Si layers for a specific wavelength (see Supplementary Information in \cite{lin2022monolithically}). Furthermore, changing to an SOI substrate would reduce the DCR.

An important performance metric is the timing jitter, which is the variance in the histogram representing the time delay between the electrical response of the detector and the arrival time of the incoming photon. In our measurements, the full-width at half-maximum (FWHM) of the histogram of the relative arrival times of the photon counts was approximately 400 ns. The fluctuation can be attributed to the laser source, measurement equipment (i.e., DSO), electrical circuitry (i.e., passive quenching circuit), and the SPAD junction. The most significant contributor to the timing performance was the rise/fall time of the quenching circuit, which also set the dead-time of photon detection. Here, the fall time of the avalanche pulses was about 270 ns, mainly due to junction capacitance $C_{j}$ and load resistance $R_{L}$. Due to the slow current buildup and reset mechanisms of the passive quenching circuit, along with fluctuations in the dead-time that impact jitter, switching to an active quenching circuit can improve the timing jitter \cite{acconcia2023timing}. A quenching circuit with a lower jitter would also enable a more accurate determination of the jitter from the SPAD itself.

Unlike conventional CMOS SPADs for free-space optical input \cite{lee2015first,niclass2007single, veerappan2014substrate}, our structure employs a lateral junction in a Si mesa. This configuration enables efficient evanescent light coupling from the SiN waveguide. 
The presence of a thin lateral junction in our SPAD design, while advantageous for integration with various photonic components on our platform,  impacts the PDE and can introduce errors in photon detection due to a fraction of excited carriers drifting into the Si substrate. 
To thoroughly assess the performance and address potential limitations, we performed an analysis on the count probabilities within pulse windows during experiments conducted in the absence of an incident light, a scenario referred to as the false-positive case for the SPAD. 
The count probability was found to be 1.4\% (3.1\%) at $V_{ex}$ of 13.3\% (19.5\%), where the PDE for TE-polarized light at 488 nm was measured to be 4.2\% (7\%). The count probabilities exhibited variability across different measurements, resulting in varying false-positive error rates (e.g., lower error rates under TM-polarized light). 
These limitations can be addressed by reducing the undesired carrier drifts and DCRs by improving the junction design. This includes optimizing the doping profile and concentration of the junction, which determines the external quantum efficiency, dark current, and breakdown voltage of an APD \cite{lerach1975analytical,yanikgonul2021integrated,morgan2021waveguide}. In Yanikgonul et al. \cite{yanikgonul2021integrated}, for example, the effect of different doping profiles, namely lateral and interdigitated, and doping concentration on a waveguide end-coupled APD performance was investigated. 
Similarly, Morgan et al. \cite{morgan2021waveguide} demonstrated an improvement in the responsivity of an evanescently-coupled PD by arranging doped regions as fingers. Additionally, reducing the dark counts of the SPAD can be achieved by introducing a doped isolation layer between the junction and the substrate or using a silicon-on-insulator (SOI) substrate to reduce the volume of the detector and to prevent undesired charged carriers from reaching the active region of the SPAD \cite{lee2015first,liu2021review,novo2014illuminated,afzalian2005physical,lee2019first}. These refinements of the junction design have the potential to significantly improve the performance of the waveguide-coupled Si SPADs in the future. 
The dark count rate can be further reduced by applying excess voltages only during specific time windows that are synchronized with the incoming light pulses. This method, referred to as gated detector operation \cite{cova1996avalanche}, eliminates the need for data processing steps such as low-pass filtering and the application of time windows in the raw data to extract single photon detection events.  Gating the detector also facilitates the removal of trapped charged carriers in the junction, thereby minimizing their contribution to the DCR \cite{cova1996avalanche}. Moreover, improved circuitry reduces the frequency of spurious avalanche pulses and dark count rates. 

 In summary, we have reported a proof-of-concept realization of room-temperature waveguide-coupled Si SPADs in a monolithically integrated Si photonic platform. The DCR can be improved by optimizing the junction design and moving to a silicon-on-insulator substrate. The PDE can be improved by modifying the SiN waveguide transition to increase the coupling into the absorption region. Custom-designed active quenching circuits would reduce the timing jitter of the devices. These waveguide SPADs enable the integration of single-photon detectors for large-scale photonic circuits, which are especially of interest for quantum information processing.

\section*{METHODS}\label{methods}

\subsection*{Linear mode operation characterization setup and setting the effective optical input power}

The characterization of the SPAD in linear mode operation was performed by using a supercontinuum laser source (NKT Photonics SuperK Fianium) operating at a 78 MHz repetition rate with a narrow-band spectral filter (NKT Photonics VIS HP8) at $\lambda =$ 488 nm and $\lambda =$ 532 nm. The light coupling into and out of the chip through edge-couplers was conducted by cleaved Nufern S405-XP single-mode fibers, which were mounted on 5-axis piezo-controlled micromanipulators for accurate alignment. For the polarization control, a polarization controller (Thorlabs CPC900) was utilized in the input fiber. The output optical power was measured by a fiber-coupled detector (Newport 818-SL-L) and readout by a power meter (Newport 2936-R), which was also used to set the fiber alignment for the Geiger mode characterization measurements of the structure. A source-measure unit (Keysight B2912A Precision-SMU) was utilized for biasing the device and measuring the current via two DC electrical probes (tungsten probe tip on MPI MP40 micropositioner), which were contacted to the anode and cathode metal pads of the structure for linear mode operation characterization (I-V characteristics). For the I-V characteristics, the bias voltage was swept from zero bias to reverse bias in steps of 80 mV. The current limit was set to 10 $\mu$A to protect the junction from damage when the device enters the Geiger mode region ($|V_{bias}|>|V_{br}|$) during I-V measurement.

The portion of the photocurrent ($I_{st}$) due to the absorption of the stray light, which was not coupled into the waveguide, was estimated by repeating the measurements in which the input fiber was horizontally shifted from its most optimal position by an amount equal to mode-field diameter ($\sim$ 3 $\mu$m) so that the contribution of waveguide-coupled light to the photocurrent was minimized. Here, it was assumed that horizontal displacement of the input fiber by approximately 3 $\mu$m would not affect absorption of the stray light while effectively eliminating light coupling into the waveguide. The edge coupler loss ($\alpha_{EC}$) was taken into account and the corresponding stray light contribution, i.e., $I_{st}(1-10^{-\alpha_{EC}/10})$, was subtracted from the overall photocurrent ($I_{ph}$). Using this approach, 39$\%$ (22\%) and 51$\%$ (34\%) of $I_{ph}$ was attributed to the stray light at wavelengths of 488 nm and 532 nm in TE (TM) polarization, respectively. The resulting photocurrent, also defined as effective photocurrent ($I_{eph}$), was used in responsivity and external quantum efficiency calculations.

\subsection*{Geiger-mode operation and quenching circuit}

The I-V curve of the diode in Figures \ref{fig:LinearMode_PD_CrossSection}(e) and \ref{fig:LinearMode_PD_CrossSection}(f) shows a breakdown voltage ($V_{br}$) beyond which the current diverges and Eq.\ref{eq:ChargedCarriers} no longer holds. In this avalanche regime, the first absorbed photon triggers the generation of additional electron-hole pairs through impact ionization in the multiplication region (depletion region) of the diode. To prevent damage to the junction, this current must be quenched by an external circuit.

A resistor connected in series with the diode serves as a passive quenching circuit (PQC) by reducing the bias voltage below the breakdown voltage after the high avalanche current is generated. The SPAD, brought into its linear-mode operation by quenching, is reset for the detection of the next photon or avalanche operation by recharging the diode and external capacitance. Active quenching, another approach, is based on utilizing active electronics to sense and apply a feedback mechanism to quench the SPAD and reset its bias voltage for the next detection operation \cite{cova1996avalanche}. Due to the simplicity in its implementation, a PQC was designed and utilized to demonstrate the SPAD operation of our PN-APDs [see Fig. \ref{fig:ExperimentSetupSchematic} (b)]. 
In the PQC, the SPAD is reverse biased with a voltage ($V_A$) higher than the breakdown voltage ($V_A > V_{br}$) through a load resistance, $R_L$. An absorbed photon triggers a high self-sustaining avalanche current, which decreases the diode voltage, $V_d$, due to the voltage drop at $R_L$. Before being quenched, the SPAD current can reach up to
\begin{equation}\label{eq:upperlimitSPADcurrent}
    I_{d,max} = \frac{V_A-|V_{br}|}{R_d} = \frac{V_{ex}}{R_d},
\end{equation}
where $V_{ex}$ and $R_d$ represent the excess bias voltage and diode resistance due to the space-charge resistance of the avalanche junction and the ohmic contact, respectively \cite{cova1996avalanche}. 
Subsequently, the current decreases exponentially, with a time constant determined by the junction capacitance $C_j$, and the external capacitance $C_e$.  
The asymptotic steady-state current limit, $I_f$, is 
\begin{equation}\label{eq:steadystatelimitSPADcurrent}
    I_f = \frac{V_{ex}}{R_d+R_L} \cong \frac{V_{ex}}{R_L},
\end{equation}
with the assumption that $R_L \gg R_d$. To have proper quenching operation, $I_f$ should be lower than the latching current level of the junction, which defines the minimum value of $R_L$ \cite{cova1996avalanche, brown1986characterization}. Another resistor, $R_S$, with low resistance is connected to the anode of the SPAD to measure the output signal. In the PQC here, $R_L$ and $R_S$ were set to 500 k$\Omega$ and 50 $\Omega$, respectively [Fig. \ref{fig:ExperimentSetupSchematic} (b)]. The electrical circuit was constructed on a printed circuit board (PCB) and shielded from external noise interference. Coaxial cables were utilized to establish the connections between the measurement instrument and the PQC. After the detection of a photon and avalanche breakdown, the junction no longer operates in Geiger-mode until $C_j$ ($\ll$0.1pF) and $C_e$ (1pF) are recharged. During this time, referred to as the ``dead-time,'' the SPAD is insensitive to incoming photons.  %
\subsection*{Single-photon condition}\label{single-photonCondition}

In classical physics, a perfectly coherent light beam is modeled as an electromagnetic wave with a constant angular frequency, phase, and amplitude. Time-invariant electric field amplitude and phase imply a stable intensity and an unchanging average photon flux over time since beam intensity is proportional to the square of the amplitude. 
However, there are statistical fluctuations on the number of photons on short time scales due to the discrete nature of photons \cite{migdall2013single}. The probability, $p_n$, to find $n$ photons in a coherent light wave can be described by
\begin{equation}\label{eq:poission_eq}
    p_n = \frac{\overline{n}^n}{n!}e^{-\overline{n}} ,
\end{equation}
which is a Poisson distribution where the average photon number is $\overline{n}$. That is, Poissonian statistics can provide the model for the photon number of a photon beam from a coherent laser source. The average number of photons per pulse is given by
\begin{equation}\label{eq:averagenumber}
    \overline{n} = \frac{P_{avg}\lambda}{f_{rep}hc},
\end{equation}
where $P_{avg}$, $f_{rep}$, and $\lambda$ are the average optical power in the waveguide, repetition rate, and wavelength of the optical pulses, respectively. Thus, the single-photon condition can be created by adjusting the average optical power of the light source such that $\overline{n}<1$. Here, to serve as the required single-photon source for investigating the SPAD performance, we employed attenuated laser pulses, as further detailed in the Single photon detection subsection in the Results section and Fig. \ref{fig:ExperimentSetupSchematic}.

\subsection*{Calculation of PDE and DCR}

To characterize the PDE and DCR of the SPAD, it was operated in Geiger mode ($|V_{bias}|>|V_{br}|$) by applying bias through a source-measure unit (Keysight B2912A Precision, SMU).  
The 0.1 ms-time traces of the repeated acquisitions were collected by the DSO (Keysight InfiniiVision DSOX4054A). 
Both the SMU and the DSO were controlled by a PC so that the SPAD was biased only during acquisitions, and each time trace was captured by the DSO for statistical analysis of the detection performance. The horizontal and vertical scales of the DSO were set to 10 $\mu$s and 2 mV, respectively. 

The collected time traces were low-pass filtered ($f_{cut}=$ 3 MHz) and were subsequently time-windowed. 
The leading edge of the time windows was aligned to account for the deterministic time delay between the incoming light pulses and the junction response, which was approximately 154 ns. 
Since the fall time of the avalanche pulses was about 270 ns, the window width was set to 500 ns. 
Avalanche pulses that exceeded the noise level were counted and classified as photon ($n_{photo}$) or dark ($n_{dark}$) counts based on their temporal positions, as depicted in Fig. \ref{fig:calculationflowchart}. These counts were then averaged over a span of 500 acquisitions. 
The PDE was calculated by
\begin{equation}\label{eq:PDEcalculationEq}
    \mathrm{PDE} (\%) = \frac{n_{photo}-n_{dark}^*}{\overline{n}}\times 100\%,
\end{equation}
where $n_{dark}^*$ is the average number of dark counts anticipated within the time windows, and $\overline{n}$ is the average photon number as calculated by Eq.\ref{eq:averagenumber}. The contribution of the stray light to the PDE was taken into account and eliminated by a similar approach to that used in linear mode operation.

On the other hand, the DCR represents the average number of dark counts per unit of time. Thus, it was calculated by
\begin{equation}\label{eq:DCRcalculationEq}
    DCR = \frac{n_{dark}}{\Delta T_{acq}},
\end{equation}
where $\Delta T_{acq}$ is the time duration of each acquisition, which in this case is set to 0.1 ms.

\subsection*{Validation of the measurement approach}

The validity of the measurement and analysis methodology used to investigate the single-photon detection performance of the device was confirmed by conducting a series of repeated measurements at room temperature, while keeping the bias voltage constant and blocking the input light. 
For example, the probability of counts occurring in the pulse window during a no-light condition (\emph{false-positive}) was calculated to be 1.4$\%$ (3.1$\%$) with $V_{ex}$ of 13.3$\%$ (19.5$\%$), which is significantly less than the PDE measured under the same $V_{ex}$ conditions.
 \section*{Data Availability}
Data underlying the results presented in this article are available from the authors upon reasonable request.

\section*{Acknowledgments}

The authors thank F.D. Chen for helpful discussions, as well as A. Stalmashonak and F. Weiss for assistance with the experimental setup.

\bibliography{main_refs}

\section*{Competing interests}
The authors declare no competing interests.

\section*{Author Contributions}

J.K.S.P. conceived the initial idea. 
The layout was generated by Z.Y. and W.D.S.. Y.L. and Z.Y. performed the device simulations reported in \cite{lin2022monolithically}. 
X.L., H.C., and G.Q.L. were responsible for device fabrication. A.G. carried out the measurements with the assistance of J.N.S. and W.D.S.. J.N.S. designed the quenching circuit. A.G., W.D.S., and J.K.S.P. analyzed the data. A.G. and J.K.S.P. co-wrote the manuscript with inputs from other co-authors. All work was done under the supervision of J.K.S.P.

\end{document}